\documentstyle[aaspp4]{article}
\begin{document}

\title{Distinguishing Between CDM and MOND: \\
Predictions for the Microwave Background}
\author{Stacy S. McGaugh}
\affil{Department of Astronomy, University of Maryland,
	College Park, MD 20742-2421 \\
	e-mail: ssm@astro.umd.edu}

\begin{abstract}
Two hypothesized solutions of the mass discrepancy problem are
cold dark matter (CDM) and modified Newtonian dynamics (MOND).
The virtues and vices of these very different hypotheses are
largely disjoint, making the process of distinguishing between them
very dependent on how we weigh disparate lines of evidence.
One clear difference is the nature of the principal mass constituent
of the universe (CDM or baryons).  This difference in the baryon fraction
($f_b \approx 0.1$ vs.\ 1) should leave a distinctive signature in the
spectrum of fluctuations in the cosmic microwave background.
Here I discuss some of the signatures which should be detectable
in the near future.  The most promising appears to be the ratio of
the amplitudes of the first two peaks relative to the intervening trough.
\end{abstract}


\section{Introduction}

Central to cosmology is the resolution of the mass discrepancy problem.
In the current standard picture, the
discrepancy between observed luminous mass and inferred dynamical mass in
extragalactic systems is attributed to the presence of nonbaryonic cold
dark matter (CDM).  However, the predictions of CDM (e.g., Navarro, Frenk, \&
White\markcite{NFW} 1997) fail the precision tests
afforded by the rotation curves of low surface brightness galaxies
(McGaugh \& de Blok 1998a\markcite{MBa};
Moore et al.\markcite{MQGL} 1999).  In contrast,
the modified Newtonian dynamics (MOND) introduced by
Milgrom\markcite{M83} (1983) as an alternative to dark matter accurately
predicted the behavior of these systems well in advance of the observations
(McGaugh \& de Blok 1998b\markcite{MBb}; de Blok \& McGaugh 1998\markcite{BM};
see also Begeman, Broeils, \& Sanders 1991\markcite{BBS};
Sanders\markcite{S96} 1996; Sanders \& Verheijen\markcite{SV} 1998).

In conventional cosmology, CDM is required for two fundamental reasons.
One is that the dynamically inferred mass density of the universe greatly
exceeds that appropriate for baryons as determined from primordial
nucleosynthesis ($\Omega_m \gg \Omega_b$; e.g., Copi, Schramm, \& Turner
1995\markcite{CST}).
The other is that gravitational formation of large scale structure proceeds
slowly (as $t^{2/3}$) in an expanding universe.  It is
only possible to reach the rich amount of structure observed at
$z=0$ from the smooth ($\Delta T/T \sim 10^{-5}$) microwave background at
$z \approx 1400$ if there is a nonbaryonic component whose density fluctuations
can grow unimpeded by radiation pressure.

It does appear possible to explain these points with MOND.
MOND is an alteration
of the force law at very small acceleration scales, $a < a_0 = 1.2 \times
10^{-8}\;{\rm cm}\,{\rm s}^{-2}$.  The low acceleration scale applies in
most disk galaxies and the universe\footnote{Note that the modification is
{\bf not} on some length scale.  The predictions of MOND therefore do not
vary by many orders of magnitude from the scales of galaxies to that of the
entire universe.} as a whole.  In the context of MOND, conventional measures
of the dynamical mass density of the universe are overestimated by a factor
which depends on the typical acceleration.  Accounting for this leads to a
very low density universe with $\Omega_m \approx \Omega_b$
(McGaugh \& de Blok 1998b\markcite{MBb}; Sanders\markcite{S98} 1998).

Perhaps the simplest possible MOND universe one can consider is one in which
$a_0$ remains constant in time (Felten 1984\markcite{F84};
Sanders\markcite{S98} 1998).  In this case, the
universe does not enter the MOND regime of very low acceleration
until $1+z \approx 2.33(\Omega_m h/0.02)^{-1/3}$, or $z \approx 1.6$.  
Everything is normal at higher redshift, so conventional results like
primordial nucleosynthesis and recombination are retained.
However, small regions can enter the MOND regime at early times as
the phase transition begins
(Sanders\markcite{S98} 1998).  Once radiation releases its hold on the baryons
($z \approx 200$ in a low density universe), these regions will
behave as if they possess a large quantity of dark matter.  Consequently,
structure forms very rapidly.  Indeed, early structure formation is
another promising way to distinguish between CDM and MOND
(Sanders\markcite{S98} 1998).
For the present purpose, it suffices to realize that if the
MOND force law is operative, structure forms much more rapidly than
the Newtonian $t^{2/3}$.  It is
not necessary to have CDM for a rich amount of large scale structure to grow
from an initially smooth cosmic microwave background.

In this paper I begin to explore how anisotropies in the microwave
background might help to distinguish between CDM and MOND.  As a starting
point, for MOND I make two basic assumptions:  that $a_0$ is constant, and
the background metric is flat in the usual Robertson-Walker sense.
This is not the only possibility for a MOND universe
(Milgrom\markcite{M89} 1989).  The acceleration constant may vary with time,
and the nature of the background geometry is unclear.  Still, this seems
like the most obvious point of departure.  In essence, I am examining the
microwave background anisotropy properties of a conventional cosmology in
which all the matter is baryonic in the amount specified by primordial
nucleosynthesis, but the amplitude of the anisotropies is not constrained
to be large by the slow growth of structure.  Failure of the assumptions should
result in a more pronounced effect on the microwave background than what
I discuss below, making it easier to distinguish between CDM and MOND.

\section{The Baryon Fraction Test}

The most obvious difference between CDM and MOND in the context of the
microwave background anisotropies is the baryon fraction.
CDM is thought to outweigh ordinary baryonic matter
by a factor of $\sim 10$ (Evrard, Metzler, \& Navarro 1996\markcite{EMN}).
The precise value depends on the Hubble constant and
on the type of system examined (McGaugh \& de Blok 1998a\markcite{MBa}).
If instead MOND is the cause of the observed mass discrepancies, there is no
CDM.  The difference between a baryon fraction $f_b \approx 0.1$ and
unity should leave a distinctive imprint on the microwave background.

The main impact of varying the baryon fraction
is on the relative amplitude of the peaks in the angular power spectrum of the
microwave background (as expanded in spherical harmonics).  In general,
increasing $f_b$ increases the baryon drag, which enhances the amplitude of 
compressional (odd numbered) peaks while
suppressing rarefaction (even numbered) peaks
(Hu, Sugiyama, \& Silk 1997\markcite{HSS}).
The precise shape of the power spectrum is thus very sensitive to $f_b$.

In order to investigate this aspect of the problem, I have used CMBFAST
(Seljak \& Zaldarriaga\markcite{cmbfast} 1996)
to compute the expected microwave background power spectrum in
several representative cases.  These have reasonable baryon fractions for
each case and baryon-to-photon ratios consistent with primordial
nucleosynthesis.  Several specific cases are illustrated in Figure 1.
These have $\Omega_b = 0.01$, 0.02, and 0.03 with $\Omega_{\rm CDM} = 0.2$
or 0 (so $f_b = 0.05$, 0.1, 0.15 or 1).  Other model parameters are held
fixed ($h = 0.75$, $T_{CMB} = 2.726$ K, $Y_p = 0.24$, $N_{\nu}$ = 3), and
adiabatic initial conditions are assumed.  As a check, models with
$\Omega_{\rm CDM} = 0.3$ and 0.4 were also run with the same baryon fractions
and $H_0$ scaled to maintain the same baryon-to-photon ratio.  As expected,
these resulted in power spectra which are indistinguishable in shape.

\begin{figure}
\plotone{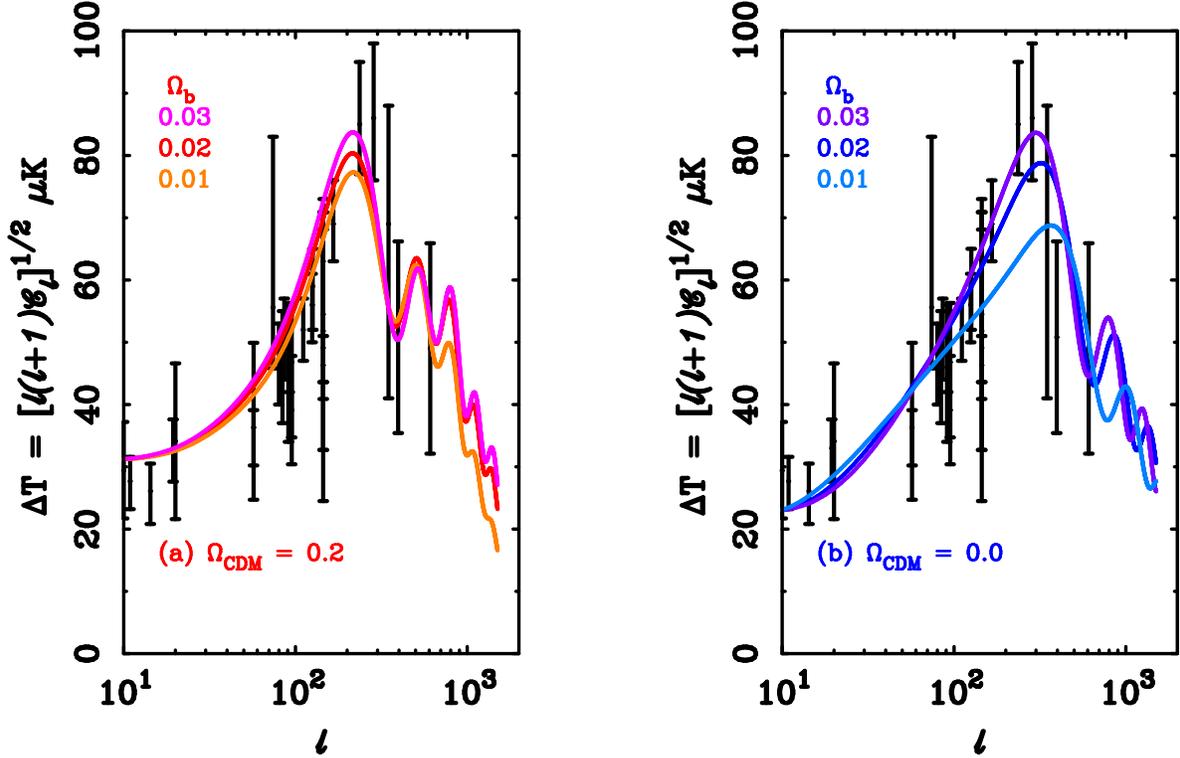}
\caption{The power spectrum of temperature anisotropies in the
microwave background with (a) and without (b) CDM.  Three choices for the
baryon density are illustrated in each case.  The highest (lowest) baryon
content corresponds to the highest (lowest) curve.  CDM models with
$\Omega_{\rm CDM} = 0.2$, 0.3, and 0.4 all gave indistinguishable results
provided the baryon fraction was the same and $H_0$ was scaled to maintain
the same baryon-to-photon ratio.  CDM models
have several distinct peaks before $\ell = 1000$ while in the pure baryon
cases representing MOND the even numbered peaks have disappeared.
Also shown are current measurements
with errors $\Delta T < 40 \mu$K from the compilation of Tegmark 
(http://www.sns.ias.edu/\~{}max/main.html\#CMB) as of March 1999.
\label{FIG1}}
\end{figure}

\placefigure{FIG1}

I am interested in the shape of the power spectrum, not the
absolute positions of the peaks.  The latter depends mostly on the scale
and geometry of the universe.  For purposes of computation, I
assume the universe is flat, with $\Omega_{\Lambda} = 1 - \Omega_{\rm CDM} -
\Omega_b$.  This results in a CDM universe close to the current
``concordant'' model (e.g., Ostriker \& Steinhardt\markcite{OS} 1995).
In the case of
MOND, the resulting model is very close to the de Sitter case.  This is
a plausible case for a MOND universe (indeed, the relation of inertial mass
to a finite vacuum energy density has been suggested as a possible physical
basis for MOND: Milgrom\markcite{M99} 1999),
but is by no means the only possibility. 
A model with no cosmological constant and $\Omega_m = \Omega_b \approx 0.02$
is plausible, but would be very open if the geometry were Robertson-Walker.
The position of the first peak in the power spectrum moves to smaller
angular scales in open universes because of the dependence of the angular
diameter distance on $\Omega_m$.  For such low $\Omega_m$ with
$\Omega_{\Lambda} = 0$,
the position of the first peak occurs at $\ell_1 > 1000$.  This is inconsistent
with recent observations which constrain $\ell_1$ to be near 200
(Miller et al.\markcite{MAT} 1999).  However, the geometry in MOND might not be
Robertson-Walker, so the position of the first peak is not uniquely specified.
It is important to realize that while the position of the first peak provides
an empirical constraint on the geometry traversed by the microwave background
photons, in the context of MOND
this does not necessarily translate into a measure of $\Omega_m$.

The test is therefore not in the absolute positions of the peaks, but in
the shape of the spectrum.  As the baryon fraction becomes
very high\footnote{Since it is possible that neutrinos have significant mass,
I also consider a model with $\Omega_{\rm CDM} = 0$ and
$\Omega_{\nu} = \Omega_b = 0.02$.  This sharpens
the peaks noticeably, but is otherwise similar to the pure baryon models.}
($f_b \rightarrow 1$), the even numbered peaks are suppressed to the point
of disappearing.  One is left with a spectrum that looks rather
like a stretched version of the standard CDM case.

The difference between the CDM and MOND cases is obvious by inspection
(Figure 1).
However, from an observer's perspective, it is not so easy to distinguish them.
The second peak has disappeared in the MOND case, so what would have been
the third peak we would now count as the second peak.  The absolute positions
of the peaks are not specified {\it a priori\/} by either theory.  The
absolute amplitude in the CDM case is constrained by the need to match
large scale structure at $z=0$.  The mechanics to do a similar exercise
with MOND do not currently exist, so the absolute amplitude is also not
specified {\it a priori}.  We must therefore rely on the relative
amplitudes and positions of the peaks to measure the
difference.  Since the third peak becomes the second peak in MOND, the
observable difference is rather more difficult to perceive than one might
have expected, at least for the assumptions made here.

The ratios of the positions and amplitudes of the peaks are given in Table 1.
The peak position ratios depend on the sound horizon at recombination,
which should not depend on MOND (for constant $a_0$) because this is well before
the universe approaches the low acceleration regime.
Other parameters do matter a bit, which can complicate matters.

One difference we could hope to distinguish is
in the ratio of the positions of the first and second peaks.  In the CDM models,
$\ell_2/\ell_1 \approx 2.35$, while in the case of MOND $\ell_2/\ell_1 \approx
2.66$.  This requires a positional accuracy determination of $\sim 5\%$
beyond $\ell > 500$, no small feat.

If we can recognize that second peak is actually missing, so that what we called
the second peak in MOND actually corresponds to the third peak in CDM, then
the distinction is greater:  for CDM, $\ell_3/\ell_1 \approx 3.6$, which
should be compared to MOND's 2.66.  It is not clear how to do this
observationally.  Once the position of the first peak is tied down, the given
ratio predicts the expected position of the second observable peak
(under the assumptions made here).  This is not very different in the two cases.

The ratios of the positions of the next observable peaks help not at all.
For CDM, $\ell_3/\ell_2 = 1.54$.  For MOND, $\ell_3/\ell_2 = 1.57$.

The ratio of the absolute amplitudes of the peaks can also distinguish the two
cases, but require comparable accuracy.  In CDM,
$(C_{{\ell},1}/C_{{\ell},2})_{abs} \approx 1.7$,
while in MOND $(C_{{\ell},1}/C_{{\ell},2})_{abs} \approx 2.4$.
This may appear to be a substantial difference, but recall that what is measured
is the temperature anisotropy.  Since $\Delta T \propto \sqrt{C_{\ell}}$,
one requires $\sim 7\%$ accuracy to distinguish the two cases at the
$2 \sigma$ level.  The
amplitude ratios of the second and third peaks have a bit more power to
distinguish between CDM and MOND, but are more difficult to measure.
The precise value of this ratio is very sensitive to $f_b$ in the CDM case.
In CDM, $(C_{{\ell},2}/C_{{\ell},3})_{abs} < 1.6$ for $f_b > 0.05$,
while in MOND $(C_{{\ell},2}/C_{{\ell},3})_{abs} \approx 1.9$.

Using the absolute amplitude of the peak heights does not untilize all the
information available.  In the purely baryonic MOND cases, there is a longer
drop from the first peak to the first trough, and a shorter rise to the
second peak than in the CDM cases.  Therefore, measuring the peak heights
relative to the bottom of the intervening trough may be a better approach.
To do this, we define $(C_{\ell,n}/C_{\ell,n+1})_{rel} = 
(C_{\ell,n} - C_{\ell,min})/(C_{\ell,n+1} - C_{\ell,min})$
to be the ratio of the amplitudes at maxima $n$ and $n+1$ less the amplitude
of the intervening minimum.  This does indeed appear more promising.
The purely baryonic MOND cases all have $(C_{\ell,1}/C_{\ell,2})_{rel} >
5$, while the CDM cases have $(C_{\ell,1}/C_{\ell,2})_{rel} < 4$
(Table 1).  This is a nice test, for in most cases this ratio
falls well on one side or the other (for $\Omega_b = 0.02$,
$(C_{\ell,1}/C_{\ell,2})_{rel}^{MOND}/(C_{\ell,1}/C_{\ell,2})_{rel}^{CDM} = 2$).

By inspection of Figure 1, one might also think that the width of the first peak
could be a discriminant, as measured at the amplitude of the first minimum.
This is a bit more sensitive to how other
parameters shift or stretch the power spectrum.  It is also very sensitive to
the neutrino mass.  Baryonic models with zero neutrino mass have perceptibly
broader peaks than the equivalent CDM model, but zero CDM models with finite
neutrino mass have peaks which are similar in width to those in the CDM models.

\section{Assumptions and Caveats}

I have made some predictions for the microwave background temperature
anisotropies which should, with sufficiently accurate measurements, distinguish
between CDM and MOND dominated universes.  The predictions are based on some
simple assumptions, most notably that the MOND acceleration constant $a_0$
does not vary substantially with time, and the geometry of the universe
is flat in the Robertson-Walker sense.  Neither of these need hold in MOND,
but plausibly may (Sanders\markcite{S98} 1998),
making this the obvious point of departure
for this discussion.  I have endeavored to make the most conservative
assumptions in the sense that failures of these assumptions should lead
to microwave background anisotropies more deviant from the standard CDM
case, and hence more readily perceptible, than the cases I have discussed.

It should be noted that the signature of a purely baryonic universe is not
necessarily reflected in the usual way in the power spectrum of large
scale structure at $z=0$ [$P(k)$ instead of $C_{\ell}$].
The calculation for the microwave background
power spectrum can be made under the assumption that MOND is not yet
important at the epoch of recombination.
It certainly is relevant by $z=0$.  The scale
which is nonlinear now is much larger in MOND than in CDM.  The rapid
nonlinear growth of structure seems likely 
to wash out the bumps and wiggles that would otherwise be imprinted on
and preserved in the power spectrum of large scale structure in the
standard framework.  So while one expects a definite signature of baryon
domination in the microwave background, one does not necessarily expect
this to be reflected in $P(k)$.

A conventional effect which may be different in the CDM and MOND
cases is reionization.  I have assumed that the background radiation
encounters effectively zero optical depth along the way to us.  However, the
optical depth can be nonzero if the universe is reionized early enough,
thus perturbing the signal in the microwave background (cf.\ Peebles \&
Jusziewicz\markcite{PJ} 1998).
Structure forms faster in MOND than in CDM, so this is a greater concern.
However, the degree to which it happens depends on the details of how stars
and other potential ionizing sources actually form, which is not
understood in either case.  The main effect of a significant
optical depth is to wash out the anisotropy signal.
This should not much perturb the observational signatures I have
discussed, which focus on the detailed structure of the peaks relative to one
another.  In purely baryonic models, it is conceivable that the amplitude
of the second peak will be amplified by this process, which formally would
invalidate the test based on the ratio of the peak-to-trough amplitudes.
However, such a microwave background power spectrum would be clearly distinct
from the standard CDM case.

The integrated Sachs-Wolfe effect is another matter which may be affected
by the rapid growth of structure in MOND.  How much depends on the unknown
details of the timing.
Matter domination does not occur in MOND until $z \approx 200$
because of the low mass density of a baryon-only universe.  Growing potentials
vary rapidly, but there is
not a tremendous amount of time between then and $z \approx 10$ when $L^*$
galaxy mass objects have collapsed (Sanders\markcite{S98} 1998).
So it is not obvious how strong this effect will be, though it can
potentially have a significant impact.

It seems unlikely that there are any effects which will cause CDM and MOND
universes to be indistinguishable once sufficiently accurate observations
of the  microwave background are obtained.  For the simple assumptions I have
made, the distinction is surprisingly subtle, but certainly present.
Any breakdown of these assumptions should lead to a greater distinction
between the two.  However, it remains a substantial challenge to
understand some of basic effects which can impact the microwave
background in the context of MOND.

\section{Conclusions}

Modern cosmological models require copious amounts of
nonbaryonic cold dark matter for well established reasons.  Yet the existence
of CDM has yet to be confirmed.
The alternative to dark matter postulated by Milgrom\markcite{M83} (1983), MOND,
has long had considerable success in describing the rotation curves of spiral
galaxies (Begeman et al.\ 1991\markcite{BBS}; Sanders\markcite{S96} 1996;
Sanders \& Verheijen\markcite{SV} 1998), a fact which has no
explanation in the standard framework.  Moreover, MOND successfully
predicted, {\it a priori}, the behavior of low surface brightness galaxies
(McGaugh \& de Blok 1998b\markcite{MBb};
de Blok \& McGaugh 1998\markcite{BM}), a test which CDM models fail
(McGaugh \& de Blok 1998a\markcite{MBa}; Moore et al.\markcite{MQGL} 1999).
Yet MOND has no clear cosmology.

In this paper, I have attempted to make some predictions for the temperature
anisotropies in the microwave background which might potentially discriminate
between CDM and MOND dominated cosmologies.  In this context,
the essential difference between the two is the baryon fraction
($f_b \approx 0.1$ for CDM and $f_b =1$ for MOND).
I have used this fact to examine the differences expected for microwave
background observations in as conservative and model independent a
way as possible.

Upcoming experiments to measure the anisotropies of the microwave background
to high precision should be able to distinguish between CDM and MOND.
For the simple assumptions investigated here,
the observational signatures are surprisingly subtle, requiring
high accuracy (i.e., peak position or amplitude to $\sim 5\%$ at $\ell > 500$.)
Perhaps the most promising test is the ratio of peak-to-trough
amplitudes of the first two peaks, with 
$(C_{\ell,1}/C_{\ell,2})_{rel} < 4$ in plausible CDM models and
$(C_{\ell,1}/C_{\ell,2})_{rel} > 5$ in MOND.

These predictions are offered in the hope of clearly distinguishing between
CDM and MOND in the near future.


\clearpage
\begin{deluxetable}{cccccc}
\tablewidth{0pt}
\tablecaption{Microwave Background Anisotropy Measures}
\tablehead{\colhead{$\Omega_{\rm CDM}$}	& \colhead{$\Omega_b$}
& \colhead{peak $n$} & \colhead{$\ell_{n+1}/\ell_n$}	&
\colhead{$(C_{\ell,n}/C_{\ell,n+1})_{abs}$} &
\colhead{$(C_{\ell,n}/C_{\ell,n+1})_{rel}$}}
\startdata
0.2	& 0.01 & 1	& 2.31	& 1.54	&	2.99 \nl
	&	& 2	& 1.55	& 1.56	&	4.80 \nl
	&	& 3	& 1.37	& 2.36 &	\nodata \nl
	& 0.02	& 1	& 2.36	& 1.60	& 2.90 \nl
	&	& 2	& 1.54 & 1.25 & 2.07 \nl
	&	& 3	& 1.38 & 2.02 & \nodata \nl
	&	& 4	& 1.26	& 1.80 & \nodata \nl
	& 0.03	& 1	& 2.40	& 1.83	& 3.48 \nl
	&	& 2	& 1.53	& 1.10	& 1.37 \nl
	&	& 3	& 1.38	& 1.97	& \nodata \nl
	&	& 4	& 1.26	& 1.61	& \nodata \nl
0.0	& 0.01	& 1	& 2.74	& 2.57	& 7.61 \nl
	& 0.02	& 1	& 2.65	& 2.37	& 5.72 \nl
	&	& 2	& 1.58	& 1.96	& 5.84 \nl
	& 0.03	& 1	& 2.62	& 2.40	& 5.41 \nl
	&	& 2	& 1.57	& 1.88	& 4.68 \nl
$\Omega_{\nu} = 0.02$ & 0.02 & 1	& 2.57	& 2.22	& 5.13 \nl
\enddata
\tablenotetext{}{Note that in the $\Omega_{\rm CDM} = 0$ models, what would
have been the even numbered peaks are completely suppressed.  The peak we call
$n=2$ for MOND corresponds to $n=3$ in CDM, and $n=3$ to $n=5$.}
\end{deluxetable}

\clearpage

\clearpage

\end{document}